\documentclass[aps,prb,groupedaddress,amsmath,eqsecnum,twocolumn]{revtex4}
    \usepackage{bm}
    \usepackage{graphicx}
    \usepackage{color}
\newcommand\beq{\begin{equation}}
\newcommand\eeq{\end{equation}}
\newcommand\beqa{\begin{eqnarray}}
\newcommand\eeqa{\end{eqnarray}}
\newcommand{\nn}{\nonumber\\}

\begin{document}
\title{Local and global properties of mixtures in one-dimensional systems. II. Exact results for the Kirkwood--Buff integrals}
\author{Arieh Ben-Naim}
\email{arieh@fh.huji.ac.il}
\affiliation{The Hebrew University of Jerusalem,
Givat Ram, Jerusalem 91904,
Israel}
\author{Andr\'es Santos}
\email{andres@unex.es}
\affiliation{Departamento de F\'{\i}sica, Universidad de
Extremadura, E-06071 Badajoz, Spain}
\date{\today}

\begin{abstract}
The Kirkwood--Buff integrals for two-component mixtures in one-dimensional
systems are calculated directly. The results are applied  to square-well particles and found to agree
with those obtained by the inversion of the Kirkwood--Buff theory of solutions.
\end{abstract}

\maketitle

\section{Introduction}
\label{sec1}
In Part I of this series\cite{BN08} one of us has studied both the global and the local properties of
mixtures of simple particles in one-dimensional system. This work has been part of a more
general advocacy in favor of the study of local properties of liquid mixtures.\cite{BN06} Instead of
the traditional study of mixtures based on the \emph{global} properties, such as excess Gibbs
energy, entropy, volume, etc. we have advocated a shift in the paradigm towards focusing
on the \emph{local} properties of the same mixtures, such as affinities between two species
(embodied in the Kirkwood--Buff integrals), and derived quantities such as local
composition, preferential solvation, and solvation thermodynamic quantities.

The local properties, though equivalent to and derivable from the global properties,
offer a host of new information on the local environments of each molecular species in the
mixture. This information is not conspicuous from the global properties. Therefore, the
study of the local quantities offer a new and more detailed and interesting view of
mixtures.

In this paper we have recalculated the Kirkwood--Buff integrals (KBI) directly for two-component mixtures of particles interacting via square-well (SW) potential.

In the next section, we outline the derivation of the pair correlation functions for two-component systems in 1D system, for arbitrary nearest-neighbor interactions.
In section \ref{sec3} we present a sample of results for
mixtures of SW particles. It is shown that the results are in quantitative
agreement with those obtained in Part I, which were based on the partition function method
and the inversion of the Kirkwood--Buff (KB) theory of solution. We have also calculated the limiting
values of the KBI when one of the species has a vanishing mole fraction, which we could not have done from the partition
function methods.

Another question examined both numerically and theoretically is the deviations from
symmetrical ideal solutions and its relation with the stability of the mixtures. It is shown
that no miscibility gap can occur in such mixtures.
\section{Theoretical background}
\label{sec2}

{It is known that the correlation and thermodynamic properties of any one-dimensional
homogeneous system in equilibrium can be derived exactly}, provided that every particle
interacts only with its nearest neighbors.\cite{SZK53,LZ71,HC04}
The aim of this section  is to present
 a self-contained  summary of the exact
solution. Although the scheme extends to any number of components, \cite{S07} here we focus on the two-component case.

\subsection{{Correlation functions}}

Let us consider a binary one-dimensional fluid mixture {at temperature $T$,  pressure $P$, and
number densities $\rho_\alpha$ ($\alpha=A,B)$. The particles are assumed to interact only between
nearest neighbors via interaction potentials $U_{\alpha\beta}(R)$. Before considering the pair correlation functions $g_{\alpha\beta}(R)$, it is convenient to introduce some probability distributions.}

 Given a particle of species
 $\alpha$ at {a certain position},   {let $p^{(\ell)}_{\alpha\beta}(R)dR$ be the conditional probability of finding as its $\ell$th neighbor in some direction a particle of species $\beta$ at a distance between $R$ and $R+dR$}.  {If $\ell\geq 2$ it is obvious that the $(\ell-1)$th neighbor of $\alpha$ in the same direction (being located at some point $R'$ between $0$ and $R$) is also a first neighbor of $\beta$. Therefore, the} following  recurrence condition holds
\beq
p^{(\ell)}_{\alpha\beta}(R)=\sum_{\gamma=A,B}\int_0^R dR' \,
p^{(\ell-1)}_{\alpha\gamma}(R')p^{(1)}_{\gamma\beta}(R-R'),
\label{n7}
\eeq
where $p_{\alpha\beta}^{(1)}(R)$ is the \emph{nearest-neighbor} probability distribution function. On physical grounds,\cite{LZ71} the ratio $p_{\alpha A}^{(1)}(R)/p_{\alpha B}^{(1)}(R)$ must become the same for $\alpha=A$ as for $\alpha=B$ in the limit of large $R$, i.e.,
\beq
\lim_{R\to\infty}\frac{p_{AA}^{(1)}(R)}{p_{AB}^{(1)}(R)}=\lim_{R\to\infty}\frac{p_{BA}^{(1)}(R)}{p_{BB}^{(1)}(R)}.
\label{x7}
\eeq
{This relation will be used later on.}
The \emph{total}  probability density of finding a particle of species $\beta$,
given that a particle of species $\alpha$ is at the origin, is
\beq
p_{\alpha\beta}(R)=\sum_{\ell=1}^\infty p_{\alpha\beta}^{(\ell)}(R).
\label{n5.2}
\eeq

The convolution structure of Eq.\ \eqref{n7} {suggests the introduction of} the Laplace transforms
\beqa
\widetilde{p}^{(\ell)}_{\alpha\beta}(s)&=&\int_0^\infty dR\, e^{-sR}p^{(\ell)}_{\alpha\beta}(R),\nn
\widetilde{p}_{\alpha\beta}(s)&=&\int_0^\infty dR\, e^{-sR}p_{\alpha\beta}(R),
\label{x1}
\eeqa
so that Eq.\ \eqref{n7} becomes
\beq
\widetilde{p}^{(\ell)}_{\alpha\beta}(s)=\sum_{\gamma=A,B}
\widetilde{p}^{(\ell-1)}_{\alpha\gamma}(s)\widetilde{p}^{(1)}_{\gamma\beta}(s).
\label{x2}
\eeq
{Equation \eqref{x2} allows us to express $\widetilde{p}^{(\ell)}_{\alpha\beta}(s)$ in terms of the nearest-neighbor distribution as}
\beq
\widetilde{\mathsf{p}}^{(\ell)}(s)=\left[\widetilde{\mathsf{p}}^{(1)}(s)\right]^\ell,
\label{n10}
\eeq
where $\widetilde{\mathsf{p}}^{(\ell)}(s)$ is the $2\times 2$ matrix of elements
 $\widetilde{p}^{(\ell)}_{\alpha\beta}(s)$.
{}From Eqs.\ \eqref{n5.2} and \eqref{n10} we get
\beqa
\widetilde{\mathsf{p}}(s)&=&{\sum_{\ell=1}^\infty \left[\widetilde{\mathsf{p}}^{(1)}(s)\right]^\ell}\nn
&=&\widetilde{\mathsf{p}}^{(1)}(s)\cdot
\left[\mathsf{I}-\widetilde{\mathsf{p}}^{(1)}(s)\right]^{-1},
\label{n12}
\eeqa
where  $\widetilde{\mathsf{p}}(s)$ is the $2\times 2$ matrix of elements
 $\widetilde{p}_{\alpha\beta}(s)$ and $\mathsf{I}$ is the $2\times 2$ unity matrix.

 Now, notice that the pair correlation function
$g_{\alpha\beta}(R)$ and the probability density ${p}_{\alpha\beta}(R)$ are simply related by ${p}_{\alpha\beta}(R)=\rho_\beta g_{\alpha\beta}(R)$ or, equivalently in Laplace space,
\beq
{\widetilde{p}_{\alpha\beta}(s)=\rho_\beta \widetilde{g}_{\alpha\beta}(s)},
\label{gab}
\eeq
{where}
\beq
{\widetilde{g}_{\alpha\beta}(s)=\int_0^\infty dR\, e^{-sR} {g}_{\alpha\beta}(R)}
\eeq
 is the Laplace transform of ${g}_{\alpha\beta}(R)$. Therefore, thanks to the one-dimensional nature of the model and the restriction to nearest-neighbor interactions, the
knowledge of the nearest-neighbor distributions
$p_{\alpha\beta}^{(1)}(R)$ suffices to obtain the pair correlation
functions $g_{\alpha\beta}(R)$. More explicitly, {from Eqs.\ \eqref{n12} and \eqref{gab} the Laplace transforms $\widetilde{g}_{\alpha\beta}(s)$ are found to be}
\beq
\widetilde{g}_{AA}(s)=\frac{1}{{\rho_T}}\frac{Q_{AA}(s)\left[1-Q_{BB}(s)\right]+Q_{AB}^2(s)}{x_A
D(s)},
\label{22}
\eeq
\beq
{\widetilde{g}_{BB}(s)=\frac{1}{{\rho_T}}\frac{Q_{BB}(s)\left[1-Q_{AA}(s)\right]+Q_{AB}^2(s)}{ x_B
D(s)}},
\label{23}
\eeq
\beq
{\widetilde{g}_{AB}(s)=\frac{1}{{\rho_T}}\frac{Q_{AB}(s)}{\sqrt{ x_Ax_B}D(s)}},
\label{24}
\eeq
where {$\rho_T=\rho_A+\rho_B$ is the total number density,} $x_\alpha=\rho_\alpha/{\rho_T}$ is the mole fraction of species $\alpha${, and we have called}
\beq
Q_{\alpha\beta}(s)\equiv\sqrt{\frac{x_\alpha}{x_\beta}} {\widetilde{p}^{(1)}_{\alpha\beta}(s)},
\label{35}
\eeq
\beq
D(s)\equiv\left[1-Q_{AA}(s)\right]\left[1-Q_{BB}(s)\right]-Q_{AB}^2(s).
\label{26}
\eeq

{The KBI in the one-dimensional case are defined by}
\beq
G_{\alpha\beta}=2\int_{0}^\infty dR\, [g_{\alpha\beta}(R)-1].
\label{KB}
\eeq
{In terms of the Laplace transform $\widetilde{g}_{\alpha\beta}(s)$, Eq.\ \eqref{KB} can be rewritten as}
\beq
{G_{\alpha\beta}=2\lim_{s\to 0}\left[\widetilde{g}_{\alpha\beta}(s)-\frac{1}{s}\right].}
\label{KB2}
\eeq

{We see that} only the nearest-neighbor distribution {${p}^{(1)}_{\alpha\beta}(R)$} is needed to close the problem.
It can be proven\cite{LZ71,HC04} that  {${p}^{(1)}_{\alpha\beta}(R)$ is just proportional to the Boltzmann factor $e^{-U_{\alpha\beta}(R)/k_BT}$ times a decaying exponential $e^{-\xi R}$, where the damping coefficient is $\xi=P/k_BT$. Therefore,}
\beq
{p_{\alpha\beta}^{(1)}(R)}=x_\beta K_{\alpha\beta}e^{-U_{\alpha\beta}(R)/k_BT}e^{-\xi R},
\label{n32b}
\eeq
{where  the proportionality constants $K_{\alpha\beta}=K_{\beta\alpha}$ (which of course depend on the thermodynamic state of the mixture) will be determined below by applying physical consistency conditions.  Taking Laplace transforms in Eq.\ \eqref{n32b} and inserting the result into Eq.\ \eqref{35} we get}
\beq
Q_{\alpha\beta}(s)=\sqrt{x_\alpha x_\beta} {K_{\alpha\beta}}\Omega_{\alpha\beta}(s+\xi),
\label{n35}
\eeq
where
\beq
{\Omega_{\alpha\beta}(s)=\int_0^\infty dR\, e^{-sR}e^{-U_{\alpha\beta}(R)/k_BT}}
\label{Omega}
\eeq
{is} the Laplace transform of
$e^{-U_{\alpha\beta}(R)/k_BT}$.

{To recapitulate, given the interaction potentials $U_{\alpha\beta}(R)$ and given a particular thermodynamic state $(P,T,x_A)$, the three correlation functions are obtained (in Laplace space) from Eqs.\ \eqref{22}--\eqref{24}, supplemented by Eqs.\ \eqref{26}, \eqref{n35}, and \eqref{Omega}.}

\subsection{{Equation of state}}

{In order to close the exact solution, it only remains to determine  the total density ${\rho_T}$ (equation of state)  and the amplitudes $K_{\alpha\beta}$ as  functions of $P$, $T$, and $x_A=1-x_B$.  As said above, they can be easily obtained by applying basic physical conditions. First, note that
Eq.\ \eqref{x7} establishes the following relationship}
\beq
{K_{AB}^2=K_{AA}K_{BB}.}
\label{x8}
\eeq
{Next, the physical condition $\lim_{R\to\infty}g_{\alpha\beta}(R)=1$ implies that $\widetilde{g}_{\alpha\beta}(s)\to 1/s$ for small $s$.
According to Eqs.\ \eqref{22}--\eqref{24}, this is only possible if $D(0)=0$, so that $D(s)\to D'(0) s$ for small $s$, where $D'(s)=dD(s)/ds$.
Thus, one has}
\beq
{[1-Q_{AA}(0)][1-Q_{BB}(0)]-Q_{AB}^2(0)=0},
\label{x3}
\eeq
\beq
{\rho_T}=\frac{Q_{AB}(0)}{\sqrt{x_Ax_B}D'(0)},
\label{x4}
\eeq
\beq
{\rho_T}=\frac{Q_{AA}(0)[1-Q_{BB}(0)]+Q_{AB}^2(0)}{x_AD'(0)},
\label{x5}
\eeq
\beq
{\rho_T}=\frac{Q_{BB}(0)[1-Q_{AA}(0)]+Q_{AB}^2(0)}{x_BD'(0)}.
\label{x6}
\eeq
  Elimination of ${\rho_T}$ {between} Eqs.\ \eqref{x4}--\eqref{x6} yields two coupled equations which, together with Eq.\ \eqref{x3}, gives
\beq
{K_{AA}=\frac{1- x_BK_{AB}\Omega_{AB}(\xi)}{ x_A\Omega_{AA}(\xi)},}
\label{33}
\eeq
\beq
{K_{BB}=\frac{1- x_AK_{AB}\Omega_{AB}(\xi)}{ x_B\Omega_{BB}(\xi)}.}
\label{34}
\eeq
{Insertion of Eqs.\ \eqref{33} and \eqref{34}} into Eq.\ \eqref{x8} allows one to obtain a quadratic equation for $K_{AB}$ whose physical root is
\beq
{K_{AB}}=\frac{1}{\Omega_{AB}(\xi)}\frac{1- \sqrt{1 -
  4 x_A x_B (1-{R})}}{2  x_A x_B{(1- R)}},
\label{KAB}
\eeq
where we have called
\beq
{R}\equiv \frac{\Omega_{AA}(\xi)\Omega_{BB}(\xi)}{\Omega_{AB}^2(\xi)}.
\label{x13}
\eeq
It is interesting to note that, since $K_{\alpha\beta}$ and $\Omega_{\alpha\beta}$ are positive definite, Eq.\ \eqref{33} and \eqref{34} imply that $x_\alpha {K_{AB}}\Omega_{AB}(\xi)<1$ for $\alpha=A,B$, i.e.,
\beq
{K_{AB}}\Omega_{AB}(\xi)<\text{min}\left(\frac{1}{x_A},\frac{1}{x_B}\right)\leq 2.
\label{x9}
\eeq
{Finally, the density ${\rho_T}$} is obtained from either of Eqs.\ \eqref{x4}--\eqref{x6}. The result  is
\begin{widetext}
\beq
{{\rho_T}(P,T,x_A)=-\frac{1}{x_A^2 K_{AA}\Omega_{AA}'(\xi)+x_B^2 K_{BB}\Omega_{BB}'(\xi)+2x_A x_B K_{AB}\Omega_{AB}'(\xi)},}
\label{rho}
\eeq
\end{widetext}
where $\Omega_{\alpha\beta}'(s)$ is the first derivative of $\Omega_{\alpha\beta}(s)$.

{Equations \eqref{33}--\eqref{x13} and \eqref{rho}  complete the full determination of $\widetilde{g}_{\alpha\beta}(s)$  and the equation of state for \emph{any} choice of the nearest-neighbor interaction potentials $U_{\alpha\beta}(x)$ and of the thermodynamic state $(P,T,x_A)$.}

\subsection{{Kirkwood--Buff integrals}\label{sec2C}}

{The KBI $G_{\alpha\beta}$ can be derived, according to Eq.\ \eqref{KB2}}, by expanding $s \widetilde{g}_{\alpha\beta}(s)$ in powers of $s$ {as $s \widetilde{g}_{\alpha\beta}(s)=1+\frac{1}{2}G_{\alpha\beta} s+\cdots$} and identifying the linear term. After some algebra one gets
\beq
G_{AB}={\rho_T} {J}+2\frac{\Omega_{AB}'(\xi)}{\Omega_{AB}(\xi)},
\label{GAB}
\eeq
\beq
G_{AA}={\rho_T} {J}-2\frac{x_B{K_{BB}}\Omega_{BB}'(\xi)}{x_A {K_{AB}}\Omega_{AB}(\xi)}-\frac{2}{{\rho_T} x_A},
\label{GAA}
\eeq
\beq
G_{BB}={\rho_T} {J}-2\frac{x_A{K_{AA}}\Omega_{AA}'(\xi)}{x_B {K_{AB}}\Omega_{AB}(\xi)}-\frac{2}{{\rho_T} x_B},
\label{GBB}
\eeq
where
\beqa
{J}&{\equiv}&x_A^2 K_{AA} \Omega_{AA}''(\xi) +
 x_B^2 K_{BB} \Omega_{BB}''(\xi)\nn
 && +
 2 x_A x_B K_{AB} \Omega_{AB}''(\xi) \nn
 &&
{ -2 x_A x_B  K_{AB}\frac{
  \Omega_{AA}'(\xi) \Omega_{BB}'(\xi) -
   \left[\Omega_{AB}'(\xi)\right]^2}{\Omega_{AB}(\xi)}.}\nn
   &&
 \label{J}
 \eeqa

The knowledge of the {KBI} allows us to obtain the (reduced) isothermal compressibility
\beq
\chi=k_BT \left(\frac{\partial{\rho_T}}{\partial P}\right)_{T, x_A}
\label{chi}
\eeq
by means of
\beqa
\chi&=&\frac{1}{1+{\rho_T} x_A x_B \Delta_{AB}}\left[1+{\rho_T}\left(x_AG_{AA}+x_B G_{BB}\right)\right.\nn
&&\left.+{\rho_T^2} x_A x_B \left(G_{AA}G_{BB}-G_{AB}^2\right)\right]
,
\label{chi2}
\eeqa
where
\beq
\Delta_{AB}\equiv G_{AA}+G_{BB}-2G_{AB}.
\label{x10}
\eeq
It can be checked that the resulting expression of $\chi$ (which, due to its length,  will be omitted here) coincides with the one obtained  as $\chi=\left(\partial{\rho_T}/\partial\xi\right)_{T,x_A}$
 from Eq.\ \eqref{rho}. This confirms the exact character of the solution.

Making use of Eqs.\ \eqref{rho}--\eqref{GBB}, it is easy to prove that
\beq
1+{\rho_T} x_Ax_B \Delta_{AB}=\frac{2}{{K_{AB}}\Omega_{AB}(\xi)}-1,
\label{x11}
\eeq
which, according to Eq.\ \eqref{x9}, is a positive definite quantity. More explicitly, from Eq.\ \eqref{KAB} we have
\beq
1+{\rho_T} x_Ax_B \Delta_{AB}=\sqrt{1-4x_Ax_B(1-{R})}.
\label{x12}
\eeq
Therefore, the denominator in Eq.\ \eqref{chi2} never vanishes and  the isothermal compressibility is well defined. This agrees with van Hove's classical proof\cite{vH50} that no phase transition can exist in this class of nearest-neighbor one-dimensional models.

Let us now obtain the KBI in the infinite dilution limit $x_A\to 0$. In that limit,  Eqs.\ \eqref{33}--\eqref{KAB} and \eqref{rho} become
\beq
K_{AA}=\frac{\Omega_{BB}(\xi)}{\Omega_{AB}^2(\xi)},\quad K_{BB}=\frac{1}{\Omega_{BB}(\xi)},\quad K_{AB}=\frac{1}{\Omega_{AB}(\xi)},
\label{33dil}
\eeq
\beq
{{\rho_T}=-\frac{\Omega_{BB}(\xi)}{\Omega_{BB}'(\xi)}.}
\label{rhodil}
\eeq
Analogously, from Eqs.\ \eqref{GAB}--\eqref{J} one gets
\beq
G_{AB}=-\frac{\Omega_{BB}''(\xi)}{\Omega_{BB}'(\xi)}+2\frac{\Omega_{AB}'(\xi)}{\Omega_{AB}(\xi)},
\label{GABdil}
\eeq
\beq
G_{AA}=-\frac{\Omega_{BB}''(\xi)}{\Omega_{BB}'(\xi)}+4\frac{\Omega_{AB}'(\xi)}{\Omega_{AB}(\xi)}-2\frac{\Omega_{AA}(\xi)\Omega_{BB}'(\xi)}{\Omega_{AB}^2(\xi)},
\label{GAAdil}
\eeq
\beq
G_{BB}=-\frac{\Omega_{BB}''(\xi)}{\Omega_{BB}'(\xi)}+2\frac{\Omega_{BB}'(\xi)}{\Omega_{BB}(\xi)},
\label{GBBdil}
\eeq
\beq
\Delta_{AB}=2\Omega_{BB}'(\xi)\left[\frac{1}{\Omega_{BB}(\xi)}-\frac{\Omega_{AA}(\xi)}{\Omega_{AB}(\xi)}\right].
\eeq
Note that special care is needed to obtain $K_{AA}$ and $G_{AA}$.

\begin{figure}[h]
\includegraphics[width=.9\columnwidth]{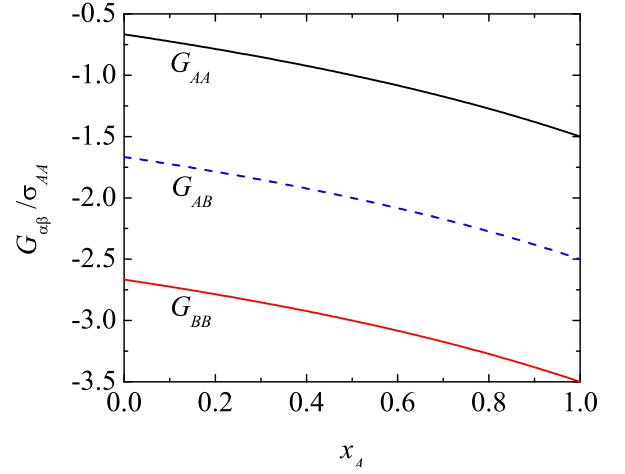}
\caption{The KBI $G_{\alpha\beta}$ for hard rods of different diameters $\sigma_{BB}/\sigma_{AA}=2$ and $P\sigma_{AA}/k_BT=1$.
\label{fig1}}
\end{figure}

\begin{figure}[h]
\includegraphics[width=.9\columnwidth]{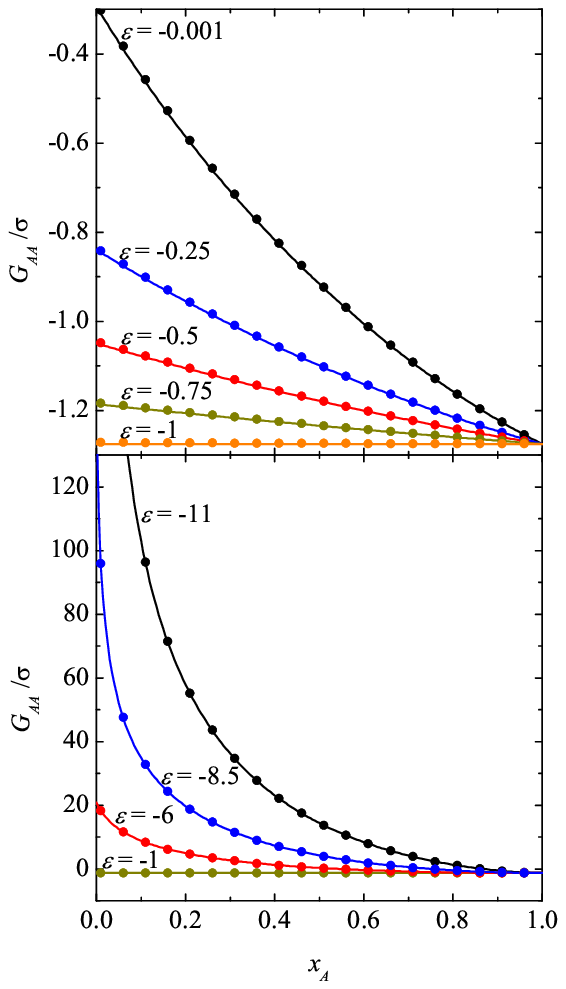}
\caption{The KBI $G_{AA}$ for SW particles with parameters given in \protect\eqref{3.3} and $k_BT/|\epsilon_{AA}| = 1$, $P\sigma/k_BT = 1$.
The lines are obtained from the exact expressions presented in Sec.\ \protect\ref{sec2C}, while the circles are the data obtained in Ref.\ \protect\onlinecite{BN08}.
\label{fig2}}
\end{figure}

\begin{figure}[h]
\includegraphics[width=.9\columnwidth]{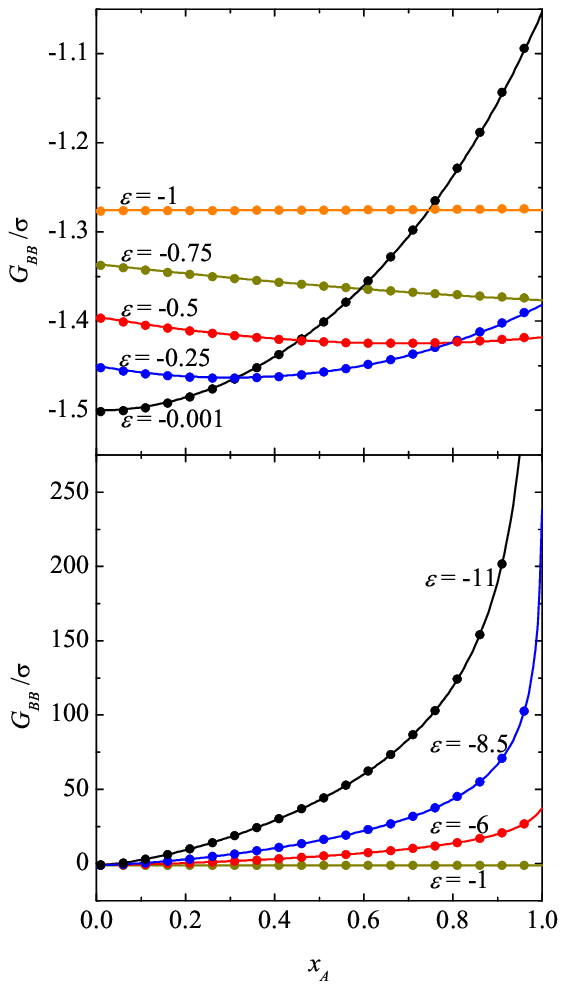}
\caption{The KBI $G_{BB}$ for SW particles with parameters given in \protect\eqref{3.3} and $k_BT/|\epsilon_{AA}| = 1$, $P\sigma/k_BT = 1$.
The lines are obtained from the exact expressions presented in Sec.\ \protect\ref{sec2C}, while the circles are the data obtained in Ref.\ \protect\onlinecite{BN08}.
\label{fig3}}
\end{figure}

\begin{figure}[h]
\includegraphics[width=.9\columnwidth]{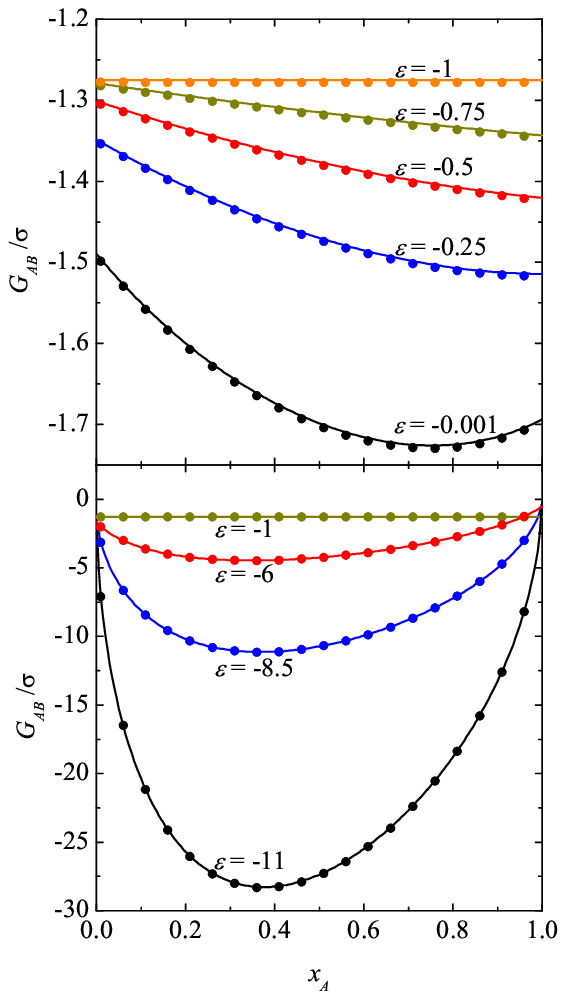}
\caption{The KBI $G_{AB}$ for SW particles with parameters given in \protect\eqref{3.3} and $k_BT/|\epsilon_{AA}| = 1$, $P\sigma/k_BT= 1$.
The lines are obtained from the exact expressions presented in Sec.\ \protect\ref{sec2C}, while the circles are the data obtained in Ref.\ \protect\onlinecite{BN08}.
\label{fig4}}
\end{figure}

\begin{figure}[h]
\includegraphics[width=.9\columnwidth]{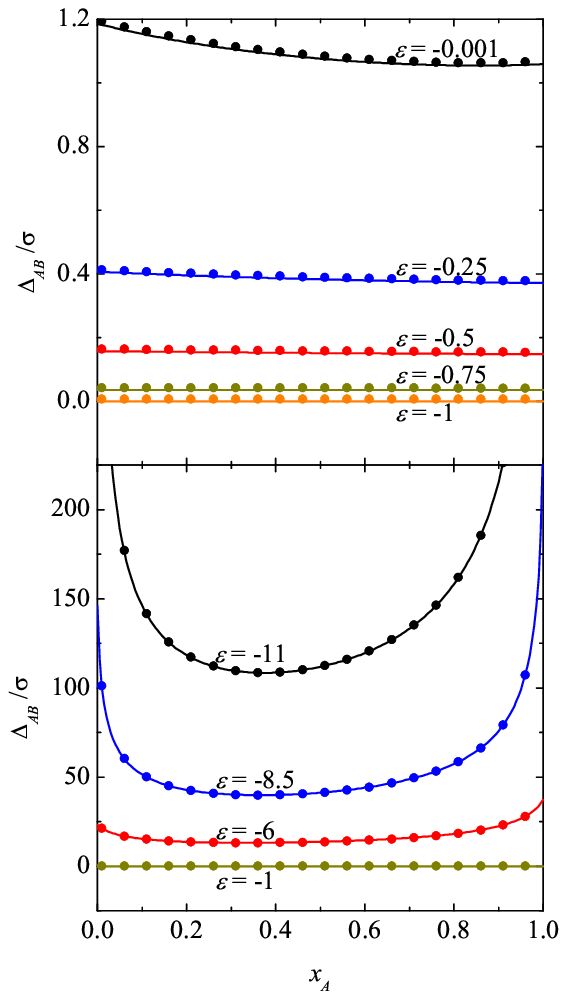}
\caption{Values of $\Delta_{AB}$ for SW particles with parameters given in \protect\eqref{3.3} and $k_BT/|\epsilon_{AA}| = 1$, $P\sigma/k_BT = 1$.
The lines are obtained from the exact expressions presented in Sec.\ \protect\ref{sec2C}, while the circles are the data obtained in Ref.\ \protect\onlinecite{BN08}.
\label{fig5}}
\end{figure}

\begin{figure}[h]
\includegraphics[width=.9\columnwidth]{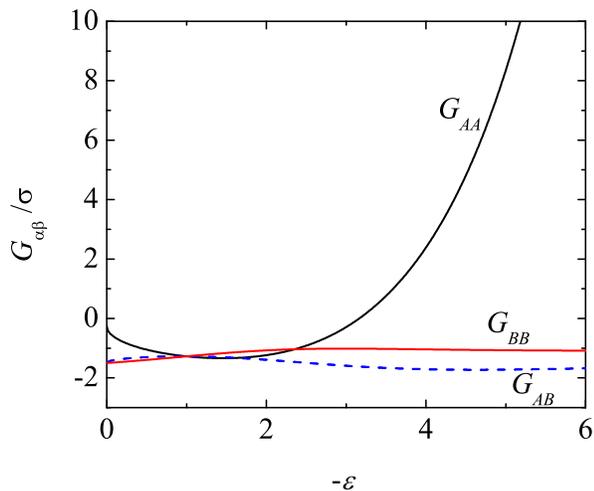}
\caption{The KBI $G_{\alpha\beta}$ in the infinite dilution limit ($x_A \to 0$) for SW particles with parameters given in \protect\eqref{3.3} and $k_BT/|\epsilon_{AA}| = 1$, $P\sigma/k_BT = 1$.
\label{fig6}}
\end{figure}

\subsection{{Chemical potentials and solvation Gibbs energies}}

Finally, let us get an explicit expression for the chemical potential. {From the KB theory of solution we have\cite{BN06,KB51}}
\beqa
\frac{1}{k_BT}\left(\frac{\partial\mu_A}{\partial x_A}\right)_{P,T}&=&\frac{1}{x_A}-\frac{{\rho_T} x_A\Delta_{AB}}{1+{\rho_T} x_Ax_B\Delta_{AB}}\nn
&{=}&{\frac{1}{x_A}-\frac{1}{x_B}\frac{\sqrt{1-4x_Ax_B(1-{R})}-1}{\sqrt{1-4x_Ax_B(1-{R})}}},\nn
&&
\label{x15}
\eeqa
{where in the last step we have made use of Eq.\ \eqref{x12}}.
Integration {over $x_A$} yields
\beqa
\frac{\mu_A}{k_BT}&=&\text{const}+\ln x_A+\ln\left[1-2 x_B (1-{R})\right.\nn
&&\left.+\sqrt{1-4x_Ax_B(1-{R})}\right].
\label{x14}
\eeqa
{For pure A ($x_B=0$), we have}
\beq
{\frac{\mu_A^P}{k_BT}=\text{const}+\ln 2.}
\label{x17}
\eeq
{The solvation Gibbs energy of A in pure A may be obtained from \eqref{x14} as\cite{BN06,BN09}}
\beq
{\Delta\mu_A^{*}=\mu_A-k_BT\ln(\rho_A\Lambda_A),}
\label{x18}
\eeq
{where $\Lambda_A=h/\sqrt{2\pi m_A k_BT}$ is the momentum partition function of A in one-dimensional systems. Similarly,}
\beq
{\Delta\mu_A^{*P}=\mu_A^P-k_BT\ln(\rho_A^P\Lambda_A),}
\label{x19}
\eeq
{where $\rho_A^P$ is the density of pure A at the same $T$ and $P$ as the mixture. Taking the limit $x_B\to 0$ in Eqs.\ \eqref{33} and \eqref{rho} one has}
\beq
{\rho_A^P=-\frac{\Omega_{AA}(\xi)}{\Omega_{AA}'(\xi)}.}
\label{x20a}
\eeq

{The excess solvation Gibbs energy relative to the solvation Gibbs energy in pure A is defined as}
\beq
{\Delta\Delta\mu_A^*=\Delta\mu_A^*-\Delta\mu_A^{*P}}
\label{x20}
\eeq
{This quantity
may be calculated from \eqref{x14}--\eqref{x20} with the result}
\beqa
\frac{\Delta\Delta\mu_A^*}{k_BT}&=&\ln\left[\frac{1}{2}-x_B (1-{R})+\frac{1}{2}\sqrt{1-4x_Ax_B(1-{R})}\right]\nn
&&+\ln\frac{\rho_A^P}{{\rho_T}}.
\label{x16}
\eeqa

\section{A sample of results}
\label{sec3}

Let us start considering a binary system composed of (additive) hard rods of different diameters (lengths) $\sigma_{AA}$, $\sigma_{BB}$, and $\sigma_{AB}= (\sigma_{AA}+\sigma_{BB})/2$. The Laplace function $\Omega_{\alpha\beta}(s)$ defined by Eq.\ \eqref{Omega} is
\beq
\Omega_{\alpha\beta}(s)=\frac{e^{-s\sigma_{\alpha\beta}}}{s}.
\label{3.2HS}
\eeq
In this case the parameter  defined in Eq.\ \eqref{x13} is $R=1$ and thus the limit $R\to 1$ must be taken in Eq.\ \eqref{KAB} with the result $K_{AB}=1/\Omega_{AB}(\xi)$. The general scheme of section \ref{sec2} can be used to obtain the KBI explicitly:
\beq
G_{AB}=-\frac{\sigma_{AA}+\sigma_{BB}+\xi \sigma_{AA}\sigma_{BB}}{1+\xi(x_A\sigma_{AA}+x_B\sigma_{BB})},
\label{GAB_HS}
\eeq
\beq
G_{AA}=G_{AB}+\sigma_{BB}-\sigma_{AA},
\label{GAA_HS}
\eeq
\beq
G_{BB}=G_{AB}+\sigma_{AA}-\sigma_{BB},
\label{GBB_HS}
\eeq
so that $\Delta_{AB}=0$.
Figure \ref{fig1} shows the values of $G_{\alpha\beta}$ for a diameter  ratio $\sigma_{BB}/\sigma_{AA}= 2$ and a thermodynamic state  $P\sigma_{AA}/k_BT=1$.  These results are in perfect agreement with those
calculated in Part I.\cite{BN08}

Having established that the programs give the correct results for hard rods, we next
present results for a mixture of particles' interaction via SW potential of the
form
\beq
U_{\alpha\beta}(R)=\begin{cases}
\infty,&R<\sigma_{\alpha\beta},\\
\epsilon_{\alpha\beta},&\sigma_{\alpha\beta}<R<\sigma_{\alpha\beta}+\delta_{\alpha\beta},\\
0,&R<\sigma_{\alpha\beta}+\delta_{\alpha\beta}.
\end{cases}
\label{3.1}
\eeq
where $\epsilon_{\alpha\beta}<0$. For this SW potential the Laplace function $\Omega_{\alpha\beta}(s)$  is
\beq
\Omega_{\alpha\beta}(s)=\frac{e^{-s\sigma_{\alpha\beta}}}{s}\left[e^{-\epsilon_{\alpha\beta}/k_BT}-\left(e^{-\epsilon_{\alpha\beta}/k_BT}-1\right)e^{-s\delta_{\alpha\beta}}\right]
\label{3.2}
\eeq
and again the general results of section \ref{sec2} provide the KBI explicitly.

We have taken the following values for the potential parameters:
\beqa
&&\sigma_{AA}=\sigma_{BB}=\sigma_{AB}=\sigma,\nn
&&\delta_{AA}=\delta_{BB}=\delta_{AB}=\frac{1}{5}\sigma,\label{3.3}\\
&& \frac{\epsilon_{BB}}{|\epsilon_{AA}|}=\epsilon,\quad \epsilon_{AB}=-\sqrt{\epsilon_{AA}\epsilon_{BB}}.\nonumber
\eeqa
The thermodynamic variables are $T$, $P$, and $x_A$. In all the calculations we choose $k_BT/|\epsilon_{AA}| = 1$ and $P\sigma/k_BT = 1$ to compare the present
results with those of Part I.

Figures \ref{fig2}--\ref{fig4} show the values of $G_{AA}$, $G_{BB}$, and $G_{AB}$ for these systems for various values
of $\epsilon$ ranging from $\epsilon = -0.001$ to $\epsilon = -1$, and from $\epsilon = -1$ to $\epsilon = -11$.\cite{note}
Figure \ref{fig5} shows the values of $\Delta_{AB}=G_{AA}+G_{BB}-2G_{AB}$ in the entire range of composition. In all the cases the
agreement with the results of Part I is quantitative.

The KBI in the infinite dilution limit ($x_A\to 0$), as obtained from Eqs.\ \eqref{GABdil}--\eqref{GBBdil}, are plotted in Fig.\ \ref{fig6} as functions of $-\epsilon$ for the same system as that of Figs.\ \ref{fig2}--\ref{fig5}. We observe that both $G_{AB}$ and $G_{BB}$ are hardly sensitive to the value of $\epsilon$. In contrast, the solute-solute KBI, $G_{AA}$, is strongly influenced by the solvent-solvent potential depth, increasing both for small and for large values of $|\epsilon|$. A careful inspection of the explicit expressions \eqref{GABdil}--\eqref{GBBdil} in the limit $|\epsilon|\to\infty$ shows that, while $G_{AB}$ and $G_{BB}$  tend to the \emph{same} constant value,  $G_{AA}$ diverges as $G_{AA}\sim \exp\left[\left(|\epsilon_{BB}|-{2}|\epsilon_{AB}|\right)/k_BT\right]$. This phenomenon might be relevant to the study of hydrophobic interactions, as discussed in Ref.\ \onlinecite{BN09}.

\section{Discussion and conclusion}
\label{sec4}
In Part I we  calculated all the KBI in an indirect way.\cite{BN08} We first calculated the excess
functions from the partition function of the system, then we used the inversion of the
KB theory\cite{BN06} to calculate the KBI. This lengthy procedure might have
introduced accumulated errors. Some readers of Part I have expressed doubts regarding the
reliability of the results calculated along this procedure. In fact some have also claimed
that there might be a miscibility gap which we might have missed by this indirect and
lengthy calculations.

In this paper we have repeated the calculations of the KBI directly, from the same program that
was designed to calculate the pair correlation functions in mixtures of two components in
1D system.

The agreement between the two methods was satisfying, it also lent credibility to the
inversion procedure and encouraged us to extend the calculations of the KBI for aqueous
like mixtures.\cite{BN09} We hope to report on that in the near future.

Regarding the question of miscibility gap we have shown that the inequality
\beq
1+{\rho_T} x_Ax_B \Delta_{AB}>0
\label{4.1}
\eeq
always holds in these mixtures, as shown by Eq.\ \eqref{x12}.

From the KB theory\cite{BN06,KB51} of solution we have the equation
\beq
\left(\frac{\partial^2 g}{\partial x_A^2}\right)_{P,T}=\frac{1}{x_B}\left(\frac{\partial\mu_A}{\partial x_A}\right)_{P,T}=\frac{k_BT}{x_A x_B (1+{\rho_T} x_Ax_B \Delta_{AB})},
\label{4.2}
\eeq
where $g=G/(N_A+N_B)$ is the Gibbs energy of the system per mole of mixture.
It follows from \eqref{4.1} and \eqref{4.2} that $g$ is everywhere a concave (downward) function of
$x_A$. Therefore, there exists no region of compositions where the system is not stable,
hence no phase transition in such a system.

\begin{acknowledgments}
The research of A.S. was supported by the Ministerio de Educaci\'on y Ciencia (Spain) through Grant No.\
FIS2007-60977 (partially financed by FEDER funds) and by the Junta
de Extremadura through Grant No.\ GRU09038.
\end{acknowledgments}


\begin{thebibliography}{99}
\bibitem{note}
It must be noted that the curves in Ref.\ \protect\onlinecite{BN08} labeled as $\epsilon=-2.5$, $-5$, $-7.5$, and $-10$ actually correspond to $\epsilon=-3.5$, $-6$, $-8.5$, and $-11$, respectively.
\end{thebibliography}
\end{document}